\documentstyle[11pt]{article}

\newcommand{\sect}[1]{ \section{#1} \setcounter{equation}{0} }
\newcommand{\partialslash}{\partial \! \! \! /}

\newcommand{\xslash}{x \! \! \! /}
\newcommand{\yslash}{y \! \! \! /}
\newcommand{\zslash}{z \! \! \! /}
\newcommand{\half}{\mbox{\small{$\frac{1}{2}$}}}

\newcommand{\la}{\langle}
\newcommand{\ra}{\rangle}
\newcommand{\Nf}{N_{\!f}}

\begin{document}
\title{Computation of critical exponent $\eta$ at $O(1/N^2_{\!f})$ in
quantum electrodynamics in arbitrary dimensions.}
\author{J.A. Gracey, \\ Department of Applied Mathematics and Theoretical
Physics, \\ University of Liverpool, \\ P.O. Box 147, \\ Liverpool, \\
L69 3BX, \\ United Kingdom.}
\date{}
\maketitle
\vspace{5cm}
\noindent
{\bf Abstract.} We present a detailed evaluation of $\eta$, the critical
exponent corresponding to the electron anomalous dimension, at $O(1/N^2_{\!f})$
in a large flavour expansion of QED in arbitrary dimensions in the Landau
gauge. The method involves solving the skeleton Dyson equations with dressed
propagators in the critical region of the theory. Various techniques to compute
massless two loop Feynman diagrams, which are of independent interest, are also
given.

\vspace{-17cm}
\hspace{10cm}
{\bf LTH-325}
\newpage
\sect{Introduction.}
The important properties of a renormalizable quantum field theory are contained
within the renormalization group equation from which one can, for instance,
determine how Green's functions depend on the renormalization scale. Central to
this equation are the $\beta$-function and $\gamma(g)$, the anomalous dimension
of the basic fields. Ordinarily one computes these within conventional
perturbation theory, in a particular renormalization scheme, and hence
determines the properties of the Green's functions to the same order. One
difficulty with perturbation theory is that computations at successive higher
orders become exceedingly tedious due in part to the increase in number of
graphs to be considered and the complexity of the integrals which appear. Thus
it is not possible to determine the higher order contributions to a
renormalization group analysis with ease. One approach to alleviate this
difficulty is to examine models in an approximation different from conventional
perturbation theory, such as the large $N$ expansion. In this approach the
quantity $gN$, where $g$ is the perturbative coupling constant, is held fixed
as
$N$ $\rightarrow$ $\infty$ so that one remains in the perturbative r\'{e}gime
of the model. Theories which admit such an expansion are those with an internal
symmetry and included in this class is quantum electrodynamics, (QED), with
$\Nf$ flavours of electrons which we will examine in detail in this paper.
{}From the graphical point of view the large $N$ expansion is a reordering of
perturbation theory such that chains of bubble graphs are summed first. As in
conventional perturbation theory one can renormalize Green's functions and
extract the pole structure from which one can deduce the large $N$
approximation to $\beta(g)$ and $\gamma(g)$. However, one will also run into
the same difficulties as perturbation theory, such as the appearance of
intractable integrals, which will occur at next to leading order.

This problem is overcome by an alternative approach developed in \cite{1,2} for
the $O(N)$ bosonic $\sigma$ model, which involves examining the theory at the
$d$-dimensional critical point, defined as the non-trivial zero of the
$\beta$-function, where the theory is finite and massless. Moreover, the fields
also obey asymptotic scaling, \cite{1}, where the propagator will take the
simple conformal structure $1/(x^2)^\alpha$ in coordinate space, with $\alpha$
its critical exponent. In particular, one examines the skeleton Dyson equations
of the theory, which are valid at the critical point, and derives a critical
point consistency equation which can be solved within the large $N$ expansion
for the anomalous dimension, \cite{1,2}. Since the anomalous dimension
exponent, $\eta$, of, say, a bosonic field is related to $\gamma(g)$ via $\eta$
$=$ $(d-2)$ $+$ $\gamma(g_c)$, where $g_c$ is the critical coupling, then one
can deduce the coefficients of $\gamma(g)$ to all orders in perturbation theory
within the particular large $N$ approximation made. This and similar scaling
relations emerge because the renormalization group equation takes a simplified
form at criticality. Further, the absence of a mass for the basic fields means
that one can quite straightforwardly probe the model beyond the leading order
of the conventional large $N$ approach.

Recently this technique was applied to QED with $\Nf$ flavours of electrons in
\cite{3}. There $\eta$ was computed at leading order in the Landau gauge and
so it is the purpose of this paper to present a detailed evaluation of the
$O(1/\Nf^2)$ corrections to the result of \cite{3} using some of the
techniques introduced in earlier works, \cite{1,2,4,5}, as well as developing
others for the specific case in hand. Since the calculation of $\eta$ is in
arbitrary dimensions, we will therefore not only provide additional
coefficients of $\gamma(g)$ in four dimensional perturbation theory in the
$\overline{\mbox{MS}}$ scheme but also $O(1/\Nf^2)$ corrections for the three
dimensional model, which is perturbatively super-renormalizable and currently
of interest in various problems. As far as we are aware the only previous
$O(1/\Nf^2)$ calculation in QED was carried out in \cite{6}, where $\eta$ was
computed in the Feynman gauge but precisely in three dimensions. Unlike the
arbitrary dimension analytic result we give here, whose derivation is
predominantly algebraic, the calculation of \cite{6} was partly carried out
numerically. Moreover, the techniques developed here for four dimensional QED,
will be very important when other more physically consistent theories like QCD
are solved within this critical point large $\Nf$ formalism. Although QED is
not
a consistent theory in isolation due to the occurrence of the Landau pole at
large values of the coupling this does not prevent us from accessing the
perturbative region of QED in large $\Nf$ as then $g$ is small.

Earlier large $\Nf$ analysis of QED was carried out in \cite{7,8} where the
pole structure of the electron self energy and photon electron vertex were
determined by explicitly carrying out the large $\Nf$ bubble sum in the Landau
gauge. The $O(1/\Nf)$ corrections to the $\beta$-function were determined as
well as the renormalization group function corresponding to the dependence of
the renormalized mass with the renormalization scale both in
$\overline{\mbox{MS}}$, \cite{7,8}, though the wave function renormalization
was not studied. To go beyond this leading order by explicitly computing the
next to leading order corrections to the bubble sum would be very involved and
thus it is appropriate to follow the more efficient and elegant methods of
\cite{1,2} to compute $O(1/\Nf^2)$ corrections since it turns out that there
are only two $2$-loop corrections to consider. Finally, we refer the interested
reader to previous $O(1/N^2)$ calculations in other models, ie
\cite{1,2,4,5,9}, since they will very much serve as a basis for the
calculation presented here. For instance, since QED involves fermions we will
use several results from the much more straightforward $O(1/N^2)$ calculation
of $\eta$ in the $O(N)$ Gross Neveu model, \cite{4}, which is also fermionic.
We note that the large $N$ exponents which have been derived in these other
models have all been shown to be in agreement with the appropriate
renormalization group functions to as many orders as they have been calculated
within explicit perturbation theory using dimensional regularization in the
$\overline{\mbox{MS}}$ scheme.

The paper is organised as follows. In section 2, we introduce the necessary
notation and formalism we will use and review the leading order analysis of
\cite{3}, but in coordinate space here. The formal corrections to the leading
order consistency equation for $\eta$ are discussed in section 3, where we also
deduce the renormalization required to analyse the theory in the critical
region. Section 4 deals with some of the calculational techniques required and
difficulties encountered, in explicitly computing the $O(1/\Nf^2)$ correction
graphs. We discuss their application to the photon self energy in section 5, as
well as providing some detail on the calculation of the component bosonic
graphs which arise and further techniques for their evaluation. A similar
discussion for the corrections to the electron self energy are given in section
6, whilst we provide the main result of our efforts in section 7, where $\eta$
is deduced at $O(1/\Nf^2)$. Our conclusions are contained in section 8.
Finally, we provide various appendices which either list useful identities for
general massless two loop Feynman diagrams or contain a library of the more
involved basic two loop component graphs which occurred in our analysis.
\sect{Preliminaries.}
We begin by introducing the formalism we will use in our analysis as well
as recalling the important features of QED we require. To fix notation and
conventions we will calculate with the (massless) QED lagrangian,
\begin{equation}
L ~=~ i \bar{\psi}^i \partialslash \psi^i + A_\mu \bar{\psi}^i \gamma^\mu
\psi^i - \frac{(F_{\mu \nu})^2}{4e^2} - \frac{(\partial_\mu A^\mu)^2}{2be^2}
\end{equation}
where unlike the usual formulation we have rescaled the coupling constant, $e$,
into the definition of the kinetic term of the $U(1)$ gauge field $A_\mu$ so
that the dimensional analysis of each term in (2.1) is completely analogous to
the earlier critical point treatment of other models, \cite{1,2}. Also, we have
set $F_{\mu \nu}$ $=$ $\partial_\mu A_\nu$ $-$ $\partial_\nu A_\mu$ and the
field $\psi^i$, $1$ $\leq$ $i$ $\leq$ $\Nf$, corresponds to the electron with
$\Nf$ flavours. The parameter, $b$, of (2.1) denotes the conventional parameter
which appears in a covariant gauge fixing term.

First, we recall the renormalization group functions, $\beta(g)$ and
$\gamma(g)$, of QED to the perturbative orders in which they are known in $d$
$=$ $4$ $-$ $2\epsilon$ dimensions $\epsilon$ $<$ $0$ small, where we set $g$
$=$ $(e/2\pi)^2$. Recently, $\beta(g)$ has been computed to four loops in
$\overline{\mbox{MS}}$ in \cite{10} and is
\begin{eqnarray}
\beta(g) &=& - \, 2\epsilon g + \frac{2\Nf}{3} g^2 + \frac{\Nf}{2} g^3
- \frac{\Nf(22\Nf + 9)}{144} g^4 \nonumber \\
&-& \frac{\Nf}{64} \left[ \frac{616\Nf^2}{243} + \left( \frac{416
\zeta(3)}{9} - \frac{380}{27} \right) \Nf + 23 \right] g^5 + O(g^6)
\end{eqnarray}
where we use the conventions of \cite{3,11} and $g$ is the dimensionless
coupling constant in $d$ dimensions. It is important to note that (2.2) is what
is determined as the $\beta$-function in $(4-2\epsilon)$-dimensions in
$\overline{\mbox{MS}}$ using dimensional regularization {\em prior} to setting
$\epsilon$ $=$ $0$ and that the coefficients are $d$-independent. (If they were
$d$-dependent then one would not be using $\overline{\mbox{MS}}$.) It is easy
to see that when $d$ $<$ $4$ (2.2) indicates the existence of a non-trivial
zero, $g_c$, of $\beta(g)$ which corresponds to a phase transition. To deduce
the perturbative coefficients of the $(4-2\epsilon)$-dimensional functions of
the renormalization group equation from the large $\Nf$ exponents we compute
later, the location of the critical point is given from (2.2) by
\begin{equation}
g_c ~=~ \frac{3\epsilon}{\Nf} - \frac{27\epsilon^2}{4\Nf^2}
+ \frac{99\epsilon^3}{16\Nf^2} + \frac{77\epsilon^4}{16\Nf^2}
+ O \left( \epsilon^5; \frac{1}{\Nf^3} \right)
\end{equation}
In (2.2), the coefficients of the large $\Nf$ term at each order in $g$ agree
with the explicit large $\Nf$ $\overline{\mbox{MS}}$ renormalization of the
$\beta$-function of QED carried out in \cite{A} using dimensional
regularization in $d$ $=$ $4$ $-$ $2\epsilon$ dimensions. Further, the critical
exponent $1/\nu$ $=$ $-$ $\beta^\prime(g_c)$, which in effect carries
information on the perturbative $\beta$-function has recently been calculated
using the formalism of \cite{1,2} by examining (2.1) at the $d$-dimensional
critical point, \cite{B}, and is in {\em exact} agreement with the results of
\cite{A,10}. The electron anomalous dimension, $\gamma(g)$, is also known to
several orders by restricting the QCD results of \cite{12,C} to the abelian
case and is
\begin{equation}
\gamma(g) ~=~ \frac{b}{2}g ~-~ \frac{(4\Nf +3)}{16}g^2 ~+~ \frac{(40\Nf^2
+ 54\Nf +27)}{576}g^3 ~+~ O(g^4)
\end{equation}
in a general covariant gauge. In fact it has been proved in \cite{D} that the
only $b$-dependence which appears in the electron anomalous dimension is at one
loop and therefore the coefficients of the two loop and higher terms of (2.4)
are gauge independent. Our remarks concerning the lack of $d$-dependence of the
coefficients of the powers of $g$ in (2.2) hold equally for this function
since it was also computed in $(4-2\epsilon)$-dimensions using dimensional
regularization with $\overline{\mbox{MS}}$. In an earlier work, \cite{3}, we
computed $\eta$ $=$ $\gamma(g_c)$ in the Landau gauge, $b$ $=$ $0$, and deduced
the $O(1/\Nf)$ terms of $\gamma(g)$ in the large $\Nf$ expansion to compare
with (2.4). These were in total agreement with the coefficients appearing in
(2.4) to $O(g^3)$ and subsequently we deduced all the higher order $O(1/\Nf)$
coefficients. So, for example, in this gauge, \cite{3},
\begin{eqnarray}
\gamma(g) &=& - \, \frac{(4\Nf+3)}{16}g^2 + \frac{(40\Nf^2 + 54\Nf +27)}{576}
g^3 \nonumber \\
&+& \left( \frac{35\Nf^3}{1296} + b_1 \Nf^2 + c_1 \Nf + d_1 \right) g^4
{}~+~ O(g^5)
\end{eqnarray}
where $b_1$, $c_1$ and $d_1$ are unknown. By calculating $\eta_2$, we will be
able to determine the next to leading order coefficients of $\gamma(g)$.
Essentially, expanding $\eta$ at a particular order in $1/\Nf$ in powers of
$\epsilon$ $=$ $2$ $-$ $\half d$, the coefficients of $\epsilon^n$ in $\eta$
are related to the $n$th order coefficient of $\gamma(g)$ since $g_c$ $\sim$
$3\epsilon/\Nf$ from (2.3). Moreover, since the first non-trivial terms of the
expansion of both $\eta$ at $O(1/\Nf)$ and $O(1/\Nf^2)$ in powers of $\epsilon$
begin at $O(\epsilon^2)$ they will produce information to all orders on the
gauge independent part of the renormalization group function. So in this sense
we will regard $\eta$ as gauge independent.

Having recalled the perturbative structure of the field theory in $d$ $=$ $4$
$-$ $2\epsilon$ dimensions and the equivalence of the leading order $1/\Nf$
exponents already computed with the technique of \cite{1,2} for QED we now
introduce the formalism of the method we use. First, we note the consequences
for the Green's functions as a result of the existence of a non-trivial fixed
point in $d$ $\neq$ $4$ dimensions. From a statistical physics point of view,
near a critical point physical quantities obey simple power law behaviour
where the power or critical exponent depends purely on the dimension of
spacetime and the parameters corresponding to any internal symmetry by the
universality principle. (See, for example, \cite{13}.) From the continuous
field theory point of view, one is dealing with fields which are conformal at
criticality and thus do not involve any mass. Therefore, we take the following
coordinate space forms for the asymptotic scaling forms of the propagators of
the fields of (2.1), which are consistent with Lorentz symmetry, near
criticality as, \cite{3,4,14},
\begin{eqnarray}
\psi(x) & \sim & \frac{A \xslash}{(x^2)^\alpha} \nonumber \\
A_{\mu \nu} (x) & \sim & \frac{B}{(x^2)^\beta} \left[ \eta_{\mu \nu}
+ \frac{2\beta}{(2\mu - 2\beta -1)} \frac{x_\mu x_\nu}{x^2} \right]
\end{eqnarray}
as $x^2$ $\rightarrow$ $0$. The point is that in this limit, or equivalently
$k^2$ $\rightarrow$ $\infty$, this structure is the dominant part of the
propagator which governs the renormalization and is therefore the part which is
relevant for deducing the anomalous parts of the critical exponents $\alpha$
and $\beta$ which are equivalent to the critical renormalization group
functions. We have chosen to work in the Landau gauge for the following reason.
Perturbatively, if one used any (covariant) gauge, other than the Landau gauge,
then the gauge parameter gets renormalized and thus the gauge changes.
Therefore, since the large $\Nf$ expansion is a reordering of perturbation
theory such that chains of bubbles are summed first, it is thus important that
one works in a gauge, ie the Landau gauge, which is unaffected by
renormalization effects, \cite{8}. Whilst $A_{\mu \nu}(x)$ does not appear to
take the usual form for the Landau gauge it is easy to transform (2.6) to
momentum space using a Fourier transform to observe that its structure will be
proportional to $P_{\mu \nu}(k)$ $=$ $(\eta_{\mu \nu}$ $-$ $k_\mu k_\nu/k^2)$
where $k$ is the momentum conjugate to $x$. For completeness and to fix
conventions, we note that the Fourier transform we use is \cite{1},
\begin{equation}
\frac{1}{(x^2)^\alpha} ~=~ \frac{a(\alpha)}{2^{2\alpha} \pi^\mu}
\int \frac{e^{ikx}}{(k^2)^{\mu - \alpha}}
\end{equation}
where we have set $a(\alpha)$ $=$ $\Gamma(\mu-\alpha)/\Gamma(\alpha)$ and
the dimension of spacetime, $d$, to be $d$ $=$ $2\mu$. For the most part of
this paper, we will work in coordinate space though it is straightforward
to map from one space to the other via (2.7) since the fields are massless.
The quantities $A$ and $B$ in (2.6) are the amplitudes of each field and are
independent of $x$ whilst $\alpha$ and $\beta$ are the exponents of the
respective fields. They are related to the exponents we will calculate, via,
\cite{3},
\begin{equation}
\alpha ~=~ \mu + \half \eta ~~~,~~~ \beta ~=~ 1 - \eta - \chi
\end{equation}
where $\eta$ is the electron anomalous dimension and $\chi$ is the anomalous
dimension of the electron photon vertex of (2.1). Both the latter quantities
depend only on $\mu$ and $\Nf$ and are $O(1/\Nf)$ within the large $\Nf$
expansion and will be calculated to $O(1/\Nf^2)$ and $O(1/\Nf)$ respectively
in this paper.

As a preliminary to introducing the formal equations which will be solved to
give an expression for $\eta_2$ we review briefly the leading order analysis.
Whilst this was done initially in \cite{3} that calculation was carried out in
momentum space. Although the same results will be obtained in coordinate space
as in momentum space we will mostly work throughout this paper in coordinate
space. The exponent $\eta$ is obtained by examining the skeleton Dyson
equations with dressed propagators at the critical point $g_c$, (2.3). As the
fields obey asymptotic scaling there then one can replace the propagators
constituting the Dyson equations with (2.6) which will therefore result in
equations involving $\alpha$ and $\beta$ which can be solved. The Dyson
equations which we consider here are illustrated in figs. 1 and 2 having been
truncated at $O(1/\Nf^2)$. (Ordinarily at $O(1/N^2)$ in this approach one has
in addition several three loop graphs but these vanish in QED due to Furry's
theorem.)

The quantities $\psi^{-1}$ and $A^{-1}_{\mu \nu}$ correspond to the inverse
propagators and their asymptotic scaling forms are obtained from (2.6) by first
transforming them to momentum space, using (2.7) and its derivatives, and
then inverting them via $G^{-1}G$ $=$ $1$, before transforming back to
coordinate space, \cite{1,2}. As in other critical point analyses of models
involving a gauge field, \cite{14}, the asymptotic form of its inverse
propagator is determined by restricting the inversion to be on the transverse
subspace of momentum space, since only the transverse piece is physically
relevant, \cite{15}. Following this procedure,
\begin{eqnarray}
\psi^{-1}(x) & \sim & \frac{r(\alpha-1) \xslash}{A (x^2)^{2\mu-\alpha+1}} \\
A^{-1}_{\mu \nu}(x) & \sim & \frac{m(\beta)}{B(x^2)^{2\mu-\beta}}
\left[ \eta_{\mu \nu} + \frac{2(2\mu-\beta)}{(2\beta-2\mu-1)} \frac{x_\mu
x_\nu}{x^2} \right]
\end{eqnarray}
as $x^2$ $\rightarrow$ $0$ where
\begin{equation}
r(\alpha) ~=~ \frac{\alpha a(\alpha-\mu)}{\pi^{2\mu}(\mu-\alpha) a(\alpha)}
{}~~~,~~~ m(\beta) ~=~ \frac{[4(\mu-\beta)^2-1]a(\beta-\mu)}
{4\pi^{2\mu}(\mu-\beta)^2 a(\beta)}
\end{equation}
and (2.9) was first given in \cite{4}.

Thus with (2.6), (2.9) and (2.10) and, for the moment, retaining only the
leading one loop graphs of figs. 1 and 2 the Dyson equations at criticality are
equivalent to
\begin{equation}
0 ~=~ r(\alpha-1) ~+~ \frac{2(2\mu-1)(\beta-\mu+1)z}{(2\mu-2\beta-1)}
\end{equation}
for the electron where we have factored off the common piece $\xslash$. Also,
\begin{equation}
0 ~=~ \frac{m(\beta)}{(x^2)^{2\mu-\beta}} \left[ \eta_{\mu \nu}
+ \frac{2(2\mu-\beta)}{(2\beta-2\mu-1)} \frac{x_\mu x_\nu}{x^2} \right]
{}~-~ \frac{4z\Nf}{(x^2)^{2\alpha-1}} \left( \eta_{\mu \nu} - \frac{2x_\mu
x_\nu}{x^2} \right)
\end{equation}
for the photon and we have set $z$ $=$ $A^2B$. As the transverse part of the
gauge field in momentum space is the only physically meaningful part of (2.13),
\cite{15}, we project this out by first transforming to momentum space,
multiply the resulting equation by the projection operator $P_{\mu \nu}(k)$,
before mapping back to coordinate space. Equivalently one can make the
following replacement for the longitudinal components of (2.13) in $x$-space,
valid for all exponents $\alpha$, which corresponds to this operation, ie
\begin{equation}
\frac{x_\mu x_\nu}{(x^2)^\alpha} ~~~ \longrightarrow ~~~ \frac{\eta_{\mu
\nu}}{2(\alpha-1) (x^2)^{\alpha-1}}
\end{equation}
Thus the relevant piece of (2.13) is
\begin{equation}
0 ~=~ \frac{(\mu-\beta)m(\beta)}{(2\mu-2\beta+1)} ~-~ \frac{4(\alpha-1)z\Nf}
{(2\alpha-1)}
\end{equation}
In writing down (2.15) we note that the powers of $x^2$, which cancel in the
leading order analysis of (2.12), are cancelled {\em after} projecting out
the relevant piece of (2.13).

We now have two equations, (2.12) and (2.15), involving two unknowns, $z$ and
$\eta$, and so eliminating $z$ between both the consistency equation for $\eta$
emerges, ie
\begin{equation}
0 ~=~ \frac{(\mu-\beta)(\beta-\mu+1)m(\beta)}{[4(\mu-\beta)^2 -1]}
{}~+~ \frac{2\Nf (\alpha-1)r(\alpha-1)}{(2\mu-1)(2\alpha-1)}
\end{equation}
which can be solved with $\alpha$ $=$ $\mu$ $+$ $\half\eta$ and $\beta$ $=$
$1$ to give
\begin{equation}
\eta_1(\mu) ~=~ - \, \frac{(2\mu-1)(2-\mu) \Gamma(2\mu)}{4\Gamma^2(\mu)
\Gamma(\mu+1) \Gamma(2-\mu)}
\end{equation}
where $\eta$ $=$ $\sum_{i=1}^{\infty} \eta_i/\Nf^i$ and hence from either
(2.12) or (2.15),
\begin{equation}
z_1 ~=~ - \, \frac{(2\mu-3)\Gamma(\mu+1) \Gamma(\mu)\eta_1}{4\pi^{2\mu}
(2\mu-1)(\mu-2)}
\end{equation}
which is required later. Although $z_1$ $=$ $0$ in three dimensions this does
not imply that the formalism breaks down in this dimension. First, there are
$O(1/\Nf^2)$ corrections to $z$ which ensure $A$ $\neq$ $0$ and $B$ $\neq$ $0$
and if one were to compute $z_1$ in arbitrary covariant gauge then the factor
becomes $(2\mu-3+b)$, \cite{13}. Second, in the determination of the final
results for exponents it turns out that the appearance of such $(2\mu-3)$
factors always arise with a factor $(2\mu-3)^{-1}$ and therefore there is no
difficulty in obtaining non-zero and non-singular results when the final
$d$-dimensional results are restricted to $d$ $=$ $3$. Indeed when (2.17) is
evaluated in three dimensions it agrees with the wave function renormalization
constant calculated in an explicit three dimensional large $\Nf$
renormalization of QED, \cite{16}, which is another strong check on our
analysis in addition to the already stated consistency of the
$\epsilon$-expansion of (2.17) with the leading order $O(g^3)$ terms of (2.14).
Whilst (2.17) was obtained in a similar fashion in momentum space the
coordinate space approach provides the starting point for computing the
corrections to $\eta_1$. For instance, one can expand (2.16) to the next order
in $1/\Nf$ but this, of course, neglects the higher order graphs which have to
be included.

\sect{Corrections to consistency equations.}
In this section, we derive the formal corrections to the consistency equations
to determine $\eta_2$ by including the higher order corrections to the skeleton
Dyson equations with dressed propagators. For QED there are only two such
correction graphs which are given in figs. 1 and 2. First, we formally denote
the values of the two loop integrals by $\Sigma$ for the electron self-energy
and $\Pi_{\mu \nu}$ $=$ $\eta_{\mu \nu}\Pi$ $+$ $x_\mu x_\nu\Xi/x^2$ for the
photon self energy in coordinate space. We will discuss their explicit
evaluation in subsequent sections but note that by value of the graph we mean
the result one would obtain by computing the integrals with unit amplitudes and
symmetry factors (such as minus signs for fermion loops etc) excluded.

Thus, the graphs of figs. 1 and 2 are equivalent to
\begin{equation}
0 ~=~ r(\alpha-1) + (x^2)^\chi f(\beta) z + (x^2)^{2\chi} z^2 \Sigma
\end{equation}
and
\begin{eqnarray}
0 &=& \frac{m(\beta)}{(x^2)^{2\mu-\beta}} \left[ \eta_{\mu \nu}
+ \frac{2(2\mu-\beta)}{(2\beta-2\mu-1)} \frac{x_\mu x_\nu}{x^2} \right]
- \frac{4z\Nf}{(x^2)^{2\alpha-1}} \left( \eta_{\mu \nu} - \frac{2x_\mu
x_\nu}{x^2} \right) \nonumber \\
&-& \frac{z^2\Nf}{(x^2)^{4\alpha+\beta-2\mu-2}} \left[ \eta_{\mu \nu}
\Pi + \frac{x_\mu x_\nu}{x^2} \Xi \right]
\end{eqnarray}
where $f(\beta)$ $=$ $2(2\mu-1)(\beta-\mu+1)/(2\mu-2\beta-1)$. Unlike at
leading order we cannot cancel off the powers of $x^2$ since now $\chi$ $=$
$O(1/\Nf)$ which will give $O(1/\Nf^2)$ contributions. Further, from the naive
computation of $\Sigma$, $\Pi$ and $\Xi$ it turns out that they are in fact
infinite due to divergences which arise from vertex subgraphs when $\chi$ $=$
$0$. This situation, indeed, is completely analogous to that at $O(1/N^2)$ in
other models, \cite{1,2,4,5}. As a first step in treating these infinities we
introduce a regulator, $\Delta$, by shifting the exponent of the gauge field,
$\beta$, by $\beta$ $\rightarrow$ $\beta$ $-$ $\Delta$, where $\Delta$ is an
infinitesimal quantity playing much the same role as $\epsilon$ $=$ $(4-d)/2$
does in dimensional regularization of four dimensional perturbative
calculations. With the introduction of $\Delta$ we formally define the
following quantities
\begin{equation}
\Sigma ~=~ \frac{K}{\Delta} + \Sigma^\prime ~~,~~
\Pi ~=~ \frac{P}{\Delta} + \Pi^\prime ~~,~~
\Xi ~=~ \frac{X}{\Delta} + \Xi^\prime
\end{equation}
where the prime, ${}^\prime$, denotes the completely finite parts of $\Sigma$,
$\Pi$ and $\Xi$ with respect to $\Delta$ and both the residues, $K$, $P$ and
$X$, and finite pieces are purely functions of $\mu$, $\alpha$ and $\beta$. The
poles with respect to $\Delta$ of (3.3) which will therefore appear in (3.1)
and (3.2) are removed by the vertex counterterm which is available at each
vertex in the one loop graphs of figs. 1 and 2. Thus denoting this counterterm
by $u$, which can be expanded about $u$ $=$ $1$ in the large $\Nf$ expansion,
as
\begin{equation}
u ~=~ 1 ~+~ \frac{u_1}{\Nf} ~+~ O\left( \frac{1}{\Nf^2} \right)
\end{equation}
the regularized Dyson equations therefore become,
\begin{equation}
0 ~=~ r(\alpha-1) + z u^2 (x^2)^{\chi+\Delta} f(\beta-\Delta)
+ z^2 (x^2)^{2\chi + 2\Delta} \left( \frac{K}{\Delta} + \Sigma^\prime
\right)
\end{equation}
and
\begin{eqnarray}
0 &=& \frac{m(\beta-\Delta)}{(x^2)^{2\mu-\beta+\Delta}} \left[ \eta_{\mu \nu}
+ \frac{2(2\mu-\beta+\Delta)}{(2\beta-2\mu-1-2\Delta)} \frac{x_\mu x_\nu}{x^2}
\right]  \nonumber \\
&-& \frac{4zu^2\Nf}{(x^2)^{2\alpha-1}} \left( \eta_{\mu \nu}
- \frac{2x_\mu x_\nu}{x^2} \right) \nonumber \\
&-& \frac{z^2\Nf}{(x^2)^{4\alpha+\beta-2\mu-2-\Delta}}
\left[ \frac{1}{\Delta} \! \left( P\eta_{\mu \nu} + X \frac{x_\mu x_\nu}{x^2}
\right) \! + \Pi^\prime \eta_{\mu \nu} + \Xi^\prime \frac{x_\mu x_\nu}{x^2}
\right] ~~~~
\end{eqnarray}
With (3.4) and expanding each term of the electron equation, (3.5), to
$O(1/\Nf^2)$ and the finite parts in $\Delta$, then the divergent terms at
$O(1/\Nf^2)$ with respect to $\Delta$ are
\begin{equation}
\frac{4u_1 z_1 (2\mu-1)(\beta-\mu+1)}{(2\mu-2\beta-1)} ~+~ \frac{Kz_1^2}
{\Delta}
\end{equation}
Setting (3.7) to zero gives a finite consistency equation for the electron so
that the $\Delta$ $\rightarrow$ $0$ limit can be achieved without difficulties.
Of course this choice corresponds to a minimal scheme. If one were to absorb
finite parts into $u_1$ then this would alter only the values of the amplitudes
but not the exponents, \cite{1}. The resulting finite equation, however,
contains $\ln x^2$ type terms which would otherwise spoil the analysis at
criticality when $x^2$ $\rightarrow$ $0$, \cite{1}. To avoid this difficulty,
$\chi$, which has yet to be determined is defined in such a way that the
$\ln x^2$ terms are absent. So with
\begin{equation}
\chi_1 ~=~ - \, \frac{(2\mu-2\beta-1)Kz_1}{2(2\mu-1)(\beta-\mu+1)}
\end{equation}
the finite Dyson equation at $O(1/\Nf^2)$ in the critical region becomes
\begin{eqnarray}
0 &=& r(\alpha-1) ~+~ \frac{2z(2\mu-1)(\beta-\mu+1)}{(2\mu-2\beta-1)}
\nonumber \\
&+& z^2 \Sigma^\prime ~+~ \frac{Kz^2}{(2\mu-2\beta-1)(\beta-\mu+1)}
\end{eqnarray}
Unlike in other $O(1/N^2)$ analyses, \cite{1,2,4,5}, where only the finite
parts of the higher order corrections contributed to $\eta_2$, here, at least
formally, the residue $K$ appears in (3.9). This is due to the fact that in the
non-regularized equation there is a non-zero function of $\beta$ multiplying
the term involving the counterterm, $u$. Thus when this function is expanded in
powers of $\Delta$, a finite term remains in the $\Delta$ $\rightarrow$ $0$
limit, when the linear term of $f(\beta-\Delta)$ multiplies the counterterm
$u_1$, which involves the residue $K$.

The treatment of the photon self energy corrections are somewhat similar to
those of the electron, though as at leading order we consider only that part of
(3.6) which corresponds to the transverse part in momentum space since this is
the physically important piece. Thus, using (2.14) we consider
\begin{eqnarray}
0 &=& \frac{2(\mu-\beta+\Delta)m(\beta-\Delta)}{(2\mu-2\beta+1+2\Delta)\Nf}
- \frac{8zu^2 (\alpha-1)(x^2)^{\chi+\Delta}}{(2\alpha-1)} \nonumber \\
&-& z^2 (x^2)^{2\chi+2\Delta} \left[ \Pi + \frac{\Xi}
{2(4\alpha+\beta-2\mu-2-\Delta)} \right]
\end{eqnarray}
where we note that the same counterterm as (3.5), of course, arises. Again
isolating the divergent terms with respect to $\Delta$ at $O(1/\Nf^2)$ they are
absorbed by setting
\begin{equation}
u_1 ~=~ - \, \frac{(2\alpha-1)z_1}{16(\alpha-1)} \left[ P + \frac{X}
{2(4\alpha+\beta-2\mu-2)} \right]
\end{equation}
which does not appear to be equivalent to that obtained from (3.7). For the
moment, we note that when the explicit values for $P$, $X$ and $K$ are
determined in a later section, we will observe that (3.7) and (3.11) are in
agreement. Further, $\ln x^2$ terms are removed by again defining $\chi_1$
appropriately. In this case, we set
\begin{equation}
\chi_1 ~=~ - \, \frac{(2\alpha-1)z_1}{8(\alpha-1)} \left[ P + \frac{X}
{2(4\alpha+\beta-2\mu-2)} \right]
\end{equation}
which will, of course, also agree with (3.8) giving us at least one check on
the explicit evaluation of $\Sigma$ and $\Pi_{\mu \nu}$. Consequently, the
finite Dyson equation for the photon at criticality is
\begin{eqnarray}
0 &=& \frac{2(\mu-\beta)m(\beta)}{(2\mu-2\beta+1)\Nf} - \frac{8z(\alpha-1)}
{(2\alpha-1)} \nonumber \\
&-& z^2 \left[ \Pi^\prime + \frac{\Xi^\prime}{2(2\alpha-1)}
+\frac{X}{2(2\alpha-1)^2} \right]
\end{eqnarray}
where, like (3.9), the residue of the two loop correction, $X$, will also give
a contribution. It arises in a different way to the appearance of $K$ in (3.9),
through the expansion in powers of $\Delta$ of the coefficient which multiplies
$\Xi$ after restriction to the transverse piece.

Whilst we have formally derived the finite corrections to (2.12) and (2.15) we
require the explicit values of $\Sigma$, $\Pi$ and $\Xi$, and hence $\chi_1$,
which is needed to evaluate the functions of $\beta$ in (3.9) and (3.13) to
$O(1/\Nf^2)$ in order to obtain the formal consistency equation satisfied by
$\eta_2$.

\sect{Computational tools for computing $\Sigma$ and $\Pi_{\mu\nu}$.}
Before detailing the explicit calculation of the two loop corrections we will
now review and develop the necessary techniques which will be required. First,
we recall that in solving the bosonic and supersymmetric $O(N)$ $\sigma$ models
at $O(1/N^2)$ in \cite{1,2,5,9}, extensive use was made of the technique known
as uniqueness which was first introduced in \cite{17} and subsequently used and
extended in various forms in [22-24]. It is applicable to models which involve
a $3$-vertex, where the exponents of the propagators forming the vertex are
initially arbitrary. In endeavouring to compute such a $3$-vertex in coordinate
space it turns out that the calculation cannot be completed in closed form
unless the sum of the exponents are restricted to be the dimension of
spacetime, $2\mu$, for a purely bosonic vertex with no derivative couplings,
\cite{17}. When this uniqueness criterion is satisfied then the integral can be
completed to yield the product of various propagators multiplied by a factor
dependent on the exponents of the initial vertex. We have summarized this rule
in fig. 3 where the Greek letter beside each line denotes the exponent of that
propagator and we have written the product of the three resulting propagators
graphically as a triangle. Also, $\nu(\alpha_1, \alpha_2, \alpha_3)$ $=$
$\pi^\mu \prod_{i=1}^3 a(\alpha_i)$. Further, if $\alpha$ $+$ $\beta$ $+$
$\gamma$ $=$ $2\mu$ $+$ $n$ for any positive integer $n$ then this vertex is
also integrable, yielding a product of triangle graphs. We have outlined this
rule in detail since in the $\sigma$ models its $3$-vertex exponents indeed sum
to the uniqueness value, \cite{2,5}, which meant that the technique could be
applied there. More recently, a similar rule was developed for the vertex of
the Gross Neveu model, where the basic uniqueness condition due to the presence
of fermions becomes $2\mu$ $+$ $1$, \cite{14}.

In the model we are concerned with here the relevant $3$-vertex contains a
gauge field interacting with two fermionic fields and therefore we need to
develop the analogous uniqueness integration value for this vertex. It is
illustrated graphically in fig. 4 in coordinate space, where the indices
$\mu$ and $\nu$ refer to those which appear in (2.6) and the numerator of the
integral is $(\yslash-\zslash)\gamma^\mu(\zslash-\xslash)$. We use the
convention that a fermion propagating from $x$ to $y$ has a factor
$(\xslash-\yslash)$. Repeating the analysis described to derive the result of
fig. 3 yields a more involved expression. After introducing Feynman parameters
for each propagator and completing two of the four integrations one obtains a
sum of various integrals each involving one hypergeometric function.
Essentially, the uniqueness condition emerges by choosing the individual
arguments of this function in such a way that it is equivalent to a simple
algebraic function, after which the integral can be computed and the result of
fig. 3 obtained for the purely bosonic vertex. In the case of fig. 4, it turns
out that the minimal uniqueness condition for this vertex is $\alpha_1$ $+$
$\alpha_2$ $+$ $\alpha_3$ $=$ $2\mu$ $+$ $2$, for the Landau gauge, which
results in a sum of triangle graphs analogous to the right side of fig. 3 but
with bosonic and fermionic propagators. We do not give the full graphical
expression here for the following simple reason. If one now examines the
electron photon vertex, then the value of the vertex, when the leading order
exponents are substituted, is $2\alpha$ $+$ $\beta$ $=$ $2\mu$ $+$ $1$, which
is one step from the uniqueness value unfortunately. This means that unlike the
models of \cite{2,4,5}, we do not have a direct integration rule with which to
perform our calculations.

To circumvent this difficulty a different approach will be required which we
now outline. The ideas we use, however, derive from similar difficulties in
other models. For instance, in the Gross Neveu model, \cite{4}, it was noted
that one can always rewrite a graph involving fermions in terms of its
constituent purely bosonic graphs, by explicitly computing the trace over
$\gamma$-matrices. A consequence of this is that some of the exponents of the
resulting graphs are reduced by an integer so that the vertices become equal to
the (bosonic) uniqueness value and therefore integrable from fig. 3. Whilst it
was possible to introduce methods in the Gross Neveu model, \cite{4}, where
taking the trace is not necessary, it is an important observation for the
present case. The application of this to QED will, in principle, be similar
though to derive the constituent bosonic graphs involves taking traces over a
larger number of $\gamma$-matrices. We will mention further simplifying aspects
later but note that in our manipulations we will make use only of the algebra
(A.1) which is, of course, valid in arbitrary dimensions.

Secondly, one technique which is used frequently in computing massless Feynman
integrals is that of integration by parts, \cite{2}. Indeed a recursion
relation was developed for a two loop integral in \cite{2} which was primarily
required for changing the value of the vertices by $\pm1$, to obtain two loop
graphs which had either integrable triangles or vertices. Equally we can apply
this technique to the gauge vertex of fig. 4 where the gauge field propagator
is
\begin{equation}
\frac{1}{(z^2)^{\alpha_1}} \left[ \eta_{\mu \nu}
+ \frac{2\alpha_1}{(2\mu-2\alpha_1-1)} \frac{z_\mu z_\nu}{z^2} \right]
\end{equation}
and its numerator was noted earlier. In the corresponding integral we choose to
rewrite part of the second term of (4.1) as
\begin{equation}
\frac{z_\mu}{(z^2)^{\alpha_1+1}} ~=~ - \, \frac{1}{2\alpha_1} \frac{\partial~}
{\partial z^\mu} \frac{1}{(z^2)^{\alpha_1}}
\end{equation}
so that when one integrates by parts, factors from differentiating the
denominators will combine with the numerator to reduce the number of
$\gamma$-matrices present by two in some terms. Thus carrying this out and
rearranging, one obtains the result illustrated graphically in fig. 5 for
arbitrary values of the exponents. Therefore, one can rewrite the $3$-vertex
involving a photon in Landau gauge in terms of three other graphs which have a
simpler structure. For instance, the first graph is proportional to what one
would obtain if the second term of (4.1) was ignored. In the remaining two
$3$-vertices the long dashed vertical line, with an index at its external end,
joining the point of integration, $z$, is to be understood as the propagator
$z_\mu/(z^2)^{\alpha_1}$, and we also note they involve only one
$\gamma$-matrix. The usefulness of this rule will become apparent later but we
remark that using it in a two loop graph will, for instance, alter the
structure of the propagator joining to the external ends of the original
$3$-vertex, in the case of the second and third terms of the right side of fig.
5. Therefore, two further rules are required, which are illustrated in figs. 6
and 7, and each is established by an integration by parts similar to (4.2). We
note that the numerator of the graph of the left side of fig. 6 is
$(\yslash-\zslash)\zslash(\zslash-\xslash)$, whilst that of fig. 7 is
$(\yslash - \xslash)\gamma_\mu$.

\sect{Computation of $\Pi_{\mu \nu}$.}
In this section, we outline the calculation of $\Pi_{\mu \nu}(x)$ and
concentrate on the main features. As discussed previously, since the method of
uniqueness cannot be used directly, we have to develop an algorithm which will
give rise to integrals to which one can apply this method.

First, the detailed graphical expression for $\Pi_{\mu \nu}(x)$ is given in
fig. 8, where the Greek indices at the vertices (internal and external)
correspond to the Lorentz index of the corresponding uncontracted
$\gamma$-matrix. Therefore, this two loop integral involves a trace over
eight $\gamma$-matrices. As a first step in reducing this number, we use the
integration by parts rules given in the previous section. Applying fig. 5 first
yields three graphs, one of which is the first graph of fig. 9, with the
remainder treated using fig. 6 at the opposite vertex. After a suitable
rearrangement they correspond to the final three graphs of fig. 9 where one
has, of course, to include the regularization $\Delta$.  We have given only the
graphical form of these graphs since the simple factors associated with each
can be readily deduced from figs. 5 and 6. The final graph is multiplied by the
factor,
\begin{equation}
\mbox{tr} (\xslash \gamma^\mu \xslash \gamma^\nu) ~=~ 4[2 x^\mu x^\nu
{}~-~ x^2 \eta^{\mu \nu}]
\end{equation}
For simplicity, we denote the contribution from the first graph of fig. 9, with
correct factors by $I$ and that from the others by $II$, treating each
separately.

To illustrate the type of calculations required once the integral has been
broken up into its constituents, we concentrate for the moment on the final
graph of fig. 9, ie $\la \alpha, \alpha, \alpha, \alpha, \beta-1 -\Delta \ra$,
where we have defined $\la \alpha_1, \alpha_2, \alpha_3, \alpha_4, \alpha_5
\ra$ to be the general bosonic two loop integral of fig. 10. To compute $\la
\alpha, \alpha, \alpha, \alpha, \beta-1-\Delta \ra$, we use the methods of
subtractions developed in \cite{2} and note that in the absence of the
regulator, $\Delta$, the graph is integrable using the uniqueness rule of fig.
3, since $2\alpha$ $+$ $(\beta$ $-$ $1)$ $=$ $2\mu$, as $\alpha$ $=$ $\mu$ and
$\beta$ $=$ $1$, at this approximation in large $\Nf$. Carrying out the naive
integration at $\Delta$ $=$ $0$, will yield a result proportional to
$\Gamma(0)$, which is infinite. However, when the theory is regularized by
$\Delta$, one loses uniqueness at each vertex and cannot naively apply fig. 3.
Instead, recalling that only the pole and finite pieces of $\Pi_{\mu \nu}$,
with respect to $\Delta$, are required to deduce $\eta_2$, one need only
consider the combination $(A-B)$ $+$ $B$, where $A$ $=$ $\la \alpha, \alpha,
\alpha, \alpha, \beta-1-\Delta \ra$, and $B$ is such that in the presence of
$\Delta$ it is calculable but has the same divergence structure as $A$. Thus,
$(A-B)$ is finite and can be deduced for $\Delta$ $=$ $0$, \cite{2}. A suitable
choice for $B$ is
\begin{equation}
\la \alpha, \alpha, \alpha, 0, \beta-1-\Delta \ra ~+~
\la \alpha, \alpha, 0, \alpha, \beta-1-\Delta \ra
\end{equation}
which corresponds to a sequence of integrable chains of propagators for
$\Delta$ $\neq$ $0$. The combination $(A-B)$ is obtained by completing
one integration with $\Delta$ $=$ $0$ first, before shifting each exponent of
the propagators of the resulting chain by a temporary regulator, $\delta$, so
that poles in $1/\delta$ can be seen to cancel before setting $\delta$ to
zero, \cite{2}. Thus, one obtains
\begin{eqnarray}
\la \alpha, \alpha, \alpha, \alpha, \beta-1-\Delta \ra &=&
\frac{2\pi^{2\mu}a^2(\alpha) a(\beta-1)}{\Gamma(\mu) \Delta} \left[
1 + \Delta \left( \!\!\! \frac{}{} B(\beta) - B(\alpha-1) \right. \right.
\nonumber \\
&+& \left. \left. \frac{1}{(\beta-1)} + \frac{(\beta-1)}{2
(\alpha-1)(\mu-\beta)} \right) \right]
\end{eqnarray}
where $B(x)$ $=$ $\psi(\mu-x)$ $+$ $\psi(x)$, and $\psi(x)$ is the logarithmic
derivative of the $\Gamma$-function, without specifying the leading order
values of $\alpha$ and $\beta$ as $\mu$ and $1$, respectively. This last remark
is actually crucial, since naively setting $\alpha$ $=$ $\mu$ and $\beta$ $=$
$1$ in (5.3) one immediately discovers that the formal expression appears to be
infinite which is due to the fact that the value $\alpha$ $=$ $\mu$ is the
anti-uniqueness value of a bosonic field, \cite{2}. In reality this problem is
not significant since there are cancelling infinities which arise from other
graphs comprising $\Pi_{\mu \nu}$, so that the sum of all the constituent
pieces is indeed finite upon setting $\alpha$ $=$ $\mu$ and $\beta$ $=$ $1$ in
the final expression. Indeed this cancellation will provide us with a stringent
check on our analysis and was also a feature of the much simpler Gross Neveu
model, \cite{4}, if one rewrites the $O(1/N^2)$ corrections there in terms of
its bosonic components.

For the remainder of the calculation one essentially breaks up the original
integral into similar bosonic two loop graphs and computes them using this
method of subtractions though some are not as straightforward as the one
outlined above. For instance, the first two graphs of fig. 11 can be treated
directly, except that in the second there is only one subtraction to consider,
since the divergence arises in the left vertex subgraph. Whilst the vertices of
the third graph of fig. 11 are unique, one cannot apply the subtraction
procedure directly, since the infinity structure of the naive subtraction $\la
\alpha, 0, \alpha-1, \alpha, \beta-\Delta \ra$ does not match that of the
original. Instead an alternative technique is required which involves rewriting
the integral in terms of other graphs, which can be computed by subtractions.
In \cite{2}, recursion relations for graphs of the type of fig. 10 were
developed by integration by parts which have the effect that the exponents of
the lines comprising the new integrals are adjusted by $\pm1$ relative to the
original. Whilst we were unable to make use of that rule here it is clear that
a similar rule will be needed.

Such an alternative was given in \cite{20,21}, being derived in the latter by
considering the uniqueness rule for a bosonic $3$-vertex when its value is
$2\mu$ $+$ $1$ and is illustrated in fig. 12. Applying it to the upper internal
vertex of the graph of fig. 10, for example, one obtains the recursion
relation, \cite{21},
\begin{eqnarray}
\la \alpha_1, \alpha_2, \alpha_3, \alpha_4, \alpha_5 \ra &=&
\frac{\alpha_5(\mu-\alpha_5-1)}{(\alpha_1-1)(\alpha_2-1)x^2}
\la \alpha_1-1, \alpha_2-1, \alpha_3, \alpha_4, \alpha_5 +1 \ra \nonumber \\
&+& \frac{(\alpha_1+\alpha_2-\mu-1)}{(\alpha_1-1)}
\la \alpha_1-1, \alpha_2, \alpha_3, \alpha_4, \alpha_5 \ra \nonumber \\
&+& \frac{(\alpha_1+\alpha_2-\mu-1)}{(\alpha_2-1)}
\la \alpha_1, \alpha_2-1, \alpha_3, \alpha_4, \alpha_5 \ra
\end{eqnarray}
Further, one can derive more relations by either applying fig. 12 to other
vertices or by first making use of the possible general transformations given
in the table of \cite{2}, then applying the rule before undoing the initial
transformation by applying its inverse. In appendix B, we have listed the other
recursion relations we required for this calculation, but note that we believe
this list is not exhaustive. Returning to the third graph of fig. 11 and
applying (B.10), results in the following three integrals, $\la \alpha, \alpha,
\alpha, \alpha, \beta - 1 - \Delta \ra$, $\la \alpha, \alpha-1, \alpha,
\alpha-1, \beta-\Delta \ra$ and $\la \alpha-1, \alpha, \alpha-1, \alpha,
\beta-\Delta \ra$, where the final two are equivalent under a change of
integration variable and each is calculable by the subtraction procedure.

In the first part of appendix C, we have listed the basic library of two loop
building block integrals we required to compute $\Pi_{\mu \nu}$, each expanded
to its finite part with respect to $\Delta$. However, we have not given the
values of those integrals which can be derived directly by subtractions, only
those which used the recursion relations given in appendix B since they are
more tedious to deduce. Where possible, we have calculated several of these
integrals in independent ways, using different recursion relations and so are
confident that the expressions are in fact correct.

One basic integral, $\la \alpha, \alpha, \alpha, \alpha, \beta-2-\Delta \ra$,
deserves special attention, since it does not have any uniqueness values at
$\Delta$ $=$ $0$ and also cannot be related to any other basic integral for
reasons which will become apparent. We compute it by first applying the
transformation $\leftarrow$, in the notation of \cite{2}, and then use the
recursion relation (B.12), which results in $\la \alpha, \mu-\alpha+1,
\mu-\alpha, \alpha, \mu-1-\Delta \ra$ and $\la \alpha-1, \mu-\alpha+1,
\mu-\alpha, \alpha, \mu-1-\Delta \ra$. Applying $\leftarrow$ to the former
yields $\la \alpha, \alpha, \alpha-1, \alpha, \beta-1-\Delta \ra$ of which only
the residue at the pole with respect to $\Delta$ is relevant since the
associated factor contains $\Delta$. One then applies (B.8) appropriately to
the other graph before undoing the initial $\leftarrow$ transformation. One of
the resulting two graphs can be treated as in an earlier discussion whilst the
other is $\la \alpha-1, \alpha-1, \alpha-1, \alpha-1, \beta-\Delta \ra$. We
were unable to evaluate this graph directly for arbitrary $\alpha$ and $\beta$,
though it is in fact finite, both with respect to $\Delta$ and setting $\alpha$
$=$ $\mu$ and $\beta$ $=$ $1$, and had therefore to leave it unevaluated in our
analysis until we set $\alpha$ $=$ $\mu$ and $\beta$ $=$ $1$ after all
contributions to $\Pi_{\mu \nu}$ had been summed. After this substitution we
can relate it to a known integral $ChT(1,1)$, in the notation of \cite{2}, by
applying the transformation $\rightarrow$. Thus,
\begin{equation}
\la \mu-1, \mu-1, \mu-1, \mu-1, 1 \ra ~=~ \frac{a^3(\mu-1)}{a(2\mu-3)}
\, ChT(1,1)
\end{equation}
and $ChT(1,1)$ $=$ $3\pi^{2\mu} a(2\mu-2) \Gamma(\mu-1) [\psi^\prime
(\mu-1) - \psi^\prime(1) ]$, \cite{28}. Adding all the pieces, we have
\begin{eqnarray}
\la \alpha, \alpha, \alpha, \alpha, \beta-2-\Delta \ra &=&
\frac{2\pi^{2\mu}(\mu-1)^2 a^2(\alpha) a(\beta-1)}{(\alpha-1)^2(\beta-1)
\Gamma(\mu)} \nonumber \\
&+& \frac{2(\beta+1-\mu)(\beta-\mu)(2\alpha-3)}{(\alpha-1)^3} \nonumber \\
&&\times ~ \la \alpha-1, \alpha-1, \alpha-1, \alpha-1, \beta - \Delta \ra ~~~
\end{eqnarray}
Several other basic building block integrals also involved $\la \alpha-1,
\alpha-1, \alpha-1, \alpha-1, \beta-\Delta \ra$, which have therefore been
left unevaluated in the list in appendix C.

We conclude the section by returning to the graphs of fig. 9 and briefly
mention the key points required to compute the first three aside from those
already discussed. For the remaining graphs of $II$, one occurs in the
evaluation of $I$, whilst the other can be treated by a subtraction, even
though it involves one fermion propagator. (The rules for integrating chains
of propagators involving fermions were given in \cite{4}, for example.)

Finally, to treat $I$ we have used the result
\begin{eqnarray}
\gamma^\rho \gamma^\mu \gamma^\nu \gamma^\sigma \gamma_\rho &=&
4 \eta^{\mu \sigma} \gamma^\nu ~-~ 2 \gamma^\mu \gamma^\sigma \gamma^\nu
\nonumber \\
&-& 2 \gamma^\nu \gamma^\mu \gamma^\sigma ~-~ 2(\mu-1) \gamma^\mu \gamma^\nu
\gamma^\sigma
\end{eqnarray}
which is valid for arbitrary dimensions and has been derived solely from (A.1).
It turns out that the three terms of (5.7) with three $\gamma$-matrices give
graphs, which although involve fermions, can be computed by the subtraction
procedure without having to explicitly project out the $\eta^{\mu \nu}$ and
$x^\mu x^\nu$ components. For instance, the graph involving only the three
$\gamma$-matrices of the last term of (5.7) is
\begin{equation}
\frac{8\pi^{2\mu}a^2(\alpha-1)a(\beta)Q_{\mu\nu}}{\Delta (\alpha-1)^2\mu
\Gamma(\mu-1)} \left[ 1 + \Delta \left( B(\beta) - B(\alpha-1)
- \frac{1}{(\alpha-1)} \right) \right]
\end{equation}
where $Q_{\mu\nu}(x)$ $=$ $\eta_{\mu\nu}$ $-$ $2x_\mu x_\nu/x^2$. For the graph
corresponding to the first term of (5.7), we had to treat its transverse and
longitudinal components separately but in doing so each piece involved at most
four $\gamma$-matrices, and therefore we employed (A.5).

Thus having given an extensive discussion of how to compute $\Pi_{\mu \nu}$ we
note that adding all contributions, $I$ is, at $\alpha$ $=$ $\mu$ and $\beta$
$=$ $1$,
\begin{equation}
- \, \frac{32 \pi^{2\mu}(\mu-1)^2 Q_{\mu\nu}}{\Delta\Gamma^2(\mu)(2\mu-3)}
\left[ \frac{1}{\mu} + \Delta \left( \frac{3}{2(\mu-1)^3}
- \frac{2}{\mu(2\mu-3)} - \frac{3\hat{\Theta}(\mu)}{2(\mu-1)} \right) \right]
\end{equation}
where $\hat{\Theta}(\mu)$ $=$ $\psi^\prime(\mu-1)$ $-$ $\psi^\prime(1)$. Whilst
we find
\begin{equation}
\frac{16\pi^{2\mu}Q_{\mu\nu}}{(2\mu-3)\Gamma^2(\mu)\Delta}
\left[ 1 - \frac{2\Delta}{(2\mu-3)} \right]
\end{equation}
for $II$. Thus summing (5.8) and (5.9) we obtain
\begin{eqnarray}
\Pi_{\mu \nu} &=& \frac{16\pi^{2\mu}}{\Delta\Gamma^2(\mu) (2\mu-3)}
\left( \eta_{\mu \nu} - \frac{2x_\mu x_\nu}{x^2} \right)
\left[ \frac{(2\mu-1)(2-\mu)}{\mu} \right. \nonumber \\
&+& \left. \Delta \left( 3(\mu-1)\hat{\Theta}(\mu) - \frac{3}{(\mu-1)}
+ \frac{2(2\mu-1)(\mu-2)}{(2\mu-3)\mu} \right) \right]
\end{eqnarray}
As mentioned earlier we have used the stringent check that the result for
$\Pi_{\mu \nu}$ must not be singular at $\alpha$ $=$ $\mu$ and $\beta$ $=$ $1$
when adding all the contributions from all the pieces. Also the tensor
structure is the same as at one loop and if one were to transform (5.11) in
the context of (3.6) to momentum space the transverse projector,
$P_{\mu \nu}(k)$, will emerge so that the result is gauge invariant. This
provides another useful check on the calculation since the individual
constituent graphs do not each have the necessary $Q_{\mu\nu}$ structure.

\sect{Computation of $\Sigma$.}

In the previous section, we outlined in detail the methods required to
determine $\Pi_{\mu\nu}$. The algorithm discussed there can also be used to
calculate $\Sigma$ though there are several differences which we mention here.
First, in $\Sigma$ of fig. 1 we have to compute a graph involving two gauge
fields. Again the first step is to break the graph up by using the integration
by parts rule on one of the gauge fields which will result in three graphs,
denoted by $I$, $II$ and $III$, respectively corresponding to the three terms
of fig. 5. To proceed further one applies either the rule of fig. 5 again or
that of fig. 7.

It turns out that the resulting nine graphs are much more straightforward to
analyse in terms of the basic bosonic two loop building block integrals than
for $\Pi_{\mu\nu}$ and in particular there is no graph analogous to $\la
\alpha, \alpha, \alpha, \alpha, \beta-2-\Delta \ra$. However, as the graph
$\Sigma$ will be proportional to $\xslash$ overall, where we use the convention
that the left external vertex of fig. 1 is the origin, the result of any
manipulations will yield graphs with at most three $\gamma$-matrices using the
results (A.2)-(A.4). To proceed further, one can either, if possible, use
fermion subtractions similar to the bosonic subtractions of section 5, which
were introduced in \cite{4}, or, if, for example, the naive subtraction is
insufficient to match the divergence structure of the particular graph then it
is best to take the explicit trace by first multiplying the graph by $\xslash$.
This will yield a sum of purely bosonic graphs which can be treated
individually. We have given the library of essential integrals in the latter
half of appendix C where we have again listed only those expressions which
were derived from recursion relations. The two loop graphs which are directly
obtained by the same type of subtractions discussed in the previous section are
not given since they are easy to deduce.

As an example we discuss the computation of the graph obtained after applying
the integration by parts rule of fig. 5 to each of the internal vertices, which
is equivalent to that of fig. 1, but where the second term of (4.1) is absent.
For concreteness, the numerator of this integral is
\begin{equation}
\gamma^\nu(-\yslash) \gamma^\sigma (\yslash-\zslash) \gamma_\nu
(\zslash-\xslash) \gamma_\sigma
\end{equation}
where $y$ is the location of the top vertex of integration and $z$ the lower
which involves seven $\gamma$-matrices. However, this number can be reduced by
using (A.4) several times and gives after a suitable rearrangement
\begin{equation}
4(\mu-1)[ 2\yslash[(x-y)^2-(y-z)^2-(x-z)^2] ~+~ (\mu-4)\yslash(\yslash
-\zslash)(\zslash-\xslash)]
\end{equation}
where we have made a change of variables to obtain the factor $2$ in the first
term. Whilst the final term involves three fermion propagators one observes
that its form is equivalent to the fermion self energy of the Gross Neveu
model, which was calculated in \cite{4}. If we introduce the notation $\la
\tilde{\alpha}_1, \alpha_2, \alpha_3, \alpha_4, \alpha_5 \ra$ to mean that the
propagator with exponent $\alpha_1$ corresponds to a fermion then $\la
\tilde{\alpha}, \beta-\Delta, \alpha, \beta-\Delta, \alpha-1 \ra$ can be
calculated directly by using the subtraction $\la \tilde{\alpha}, \beta-\Delta,
\alpha, 0, \alpha-1 \ra$ and the result, to the finite part in $\Delta$, is
\begin{equation}
\frac{\pi^{2\mu} a^2(\alpha-1) a(\beta)}{\Delta \Gamma(\mu) (\mu-\alpha)
(\alpha-1)} ~+~ O(\Delta)
\end{equation}
Again, one observes that this expression is infinite at $\alpha$ $=$ $\mu$ and
$\beta$ $=$ $1$ but in the overall sum for $\Sigma$, such infinities will in
fact cancel. For the remaining graphs, $\la \tilde{\alpha}, \beta - \Delta,
\alpha-1, \beta-\Delta, \alpha \ra$ and $\la \tilde{\alpha}, \beta-1-\Delta,
\alpha, \beta-\Delta, \alpha \ra$, one has to take a trace for each to obtain
\begin{equation}
\frac{2\pi^{2\mu}a(\alpha-1)a(\alpha)a(\beta)}{\Delta\Gamma(\mu)}
\left[ 1 + \Delta \left( B(\beta) - B(\alpha-1) - \frac{1}{\mu}
+\frac{(\beta-1)}{4\mu(\alpha-1)} \right) \right]
\end{equation}
for the former and
\begin{eqnarray}
&& \frac{\pi^{2\mu} a(\alpha) a(\alpha-1) a(\beta)}{2\Delta \Gamma(\mu)}
\left[ \left( 6 - \frac{(2\mu+1)(\beta-1)}{\mu(\alpha-1)} \right) \right.
\nonumber \\
&&+ \left. \Delta \left( \left( 2 - \frac{(\beta-1)}{\mu(\alpha-1)} \right)
[B(\beta) - B(\alpha-1) ] + \frac{\alpha(\beta-1)}{(\alpha-1)^2}
- \frac{4}{\mu} \right. \right. \nonumber \\
&&- \left. \left. \frac{(\beta-1)(\mu-1)}{\mu(\alpha-1)^2}
- \frac{(\beta-1)(\mu^2-3\mu+1)}{\mu^2(\alpha-1)} \right) \right]
\end{eqnarray}
for the latter. The computation of the remaining graphs is also similar.

Finally, we close this section by giving the results of the three component
graphs which comprise $\Sigma$, as an aid to the interested reader, having
checked that each is finite after the substitution of $\alpha$ $=$ $\mu$ and
$\beta$ $=$ $1$. Thus
\begin{eqnarray}
\Sigma_I &=& \frac{16\pi^{2\mu}(2\mu-1)(\mu-1)^2}{\Delta (2\mu-3)^2 \mu
\Gamma^2(\mu)} \left[ \mu-2 + \Delta \! \left( \frac{3\mu-4}{2\mu} -
\frac{4(\mu-2)}{(2\mu-3)} \! \right) \right] ~~~~ \\
\Sigma_{II} &=& - \, \frac{8\pi^{2\mu}(2\mu-1)}{\Delta (2\mu-3)^2
\Gamma^2(\mu)} \left[ \mu - 2 + \Delta \left( \frac{2}{\mu} -
\frac{4(\mu-2)}{(2\mu-3)} \right) \right] \\
\Sigma_{III} &=& - \, \frac{4\pi^{2\mu} (2\mu-1)(\mu-2)}{(2\mu-3)^2
\mu \Gamma^2(\mu)}
\end{eqnarray}
We note that one of the constituent graphs of $\Sigma_{II}$ involves the
relatively large expression (C.17) which is deduced from a recursion relation
containing (C.15) and (C.16). Hence,
\begin{eqnarray}
\Sigma &=& \frac{4\pi^{2\mu}(2\mu-1)}{\mu\Gamma^2(\mu)(2\mu-3)^2\Delta}
\left[ 2(2\mu-1)(\mu-2)^2 \right. \nonumber \\
&+&\left. \Delta \left( (2\mu-5)\mu + \frac{4(\mu-1)^2(\mu-2)}{\mu}
- \frac{8(2\mu-1)(\mu-2)^2}{(2\mu-3)} \right) \right]
\end{eqnarray}

We conclude our discussion of the two loop calculations by remarking that in
computing both $\Pi_{\mu \nu}$ and $\Sigma$ we made extensive use of the
relation $2\alpha$ $+$ $\beta$ $=$ $2\mu$ $+$ $1$ between the exponents
$\alpha$ and $\beta$, which is valid at the order we are calculating, to
simplify substantial amounts of tedious algebra.

\sect{Derivation of $\eta_2$.}

Having discussed the derivation of the two loop corrections we now return to
the formalism developed in section 3 and derive the consistency equation for
$\eta_2$. From (5.11) and (6.9), we first of all check that the values one
obtains for $\chi_1$ in both renormalizations agree. Thus with
\begin{equation}
K ~=~ \frac{8\pi^{2\mu}(2\mu-1)^2(\mu-2)^2}{(2\mu-3)^2 \Gamma(\mu)
\Gamma(\mu+1)} ~~,~~ P ~=~ \frac{16\pi^{2\mu}(2\mu-1)(2-\mu)}
{(2\mu-3)\Gamma(\mu) \Gamma(\mu+1)}
\end{equation}
then
\begin{equation}
\chi_1 ~=~ - \, \eta_1
\end{equation}
from either (3.8) or (3.12), and, as was noted in \cite{22}, this corresponds
to the QED Ward identity. Likewise, the expressions for the vertex
renormalization constant, $u$, both agree. With this value for $\chi_1$, then
$\beta$ $=$ $1$ $+$ $O(1/\Nf^2)$, so that we can now write down the formal
consistency equation for $\eta_2$, which is given by eliminating $z_2$ from
(3.9) and (3.13), as
\begin{eqnarray}
\frac{2\eta_2}{\eta^2_1} &=& \frac{1}{\mu} - \frac{1}{(2\mu-1)(\mu-1)}
\nonumber \\
&+& \frac{(2\mu-3)^2\Gamma(\mu)\Gamma(\mu+1)}{4\pi^{2\mu}(2\mu-1)^2
(\mu-2)^2} \left( \Sigma^\prime - \frac{K}{(2\mu-3)(\mu-2)} \right) \nonumber
\\
&+& \frac{(2\mu-3)\Gamma(\mu) \Gamma(\mu+1)}{16\pi^{2\mu}(\mu-1)(\mu-2)}
\left( \Pi^\prime + \frac{\Xi^\prime}{2(2\mu-1)} + \frac{X}{2(2\mu-1)^2}
\right)
\end{eqnarray}
The explicit expressions for $\Pi^\prime$ and $\Sigma^\prime$ can be read off
from (5.11) and (6.9), respectively, and with (7.1) some straightforward
algebra therefore leads to
\begin{eqnarray}
\eta_2 &=& \eta^2_1 \left[ \frac{3\mu(\mu-1)\hat{\Theta}(\mu)}{(2\mu-1)
(\mu-2)} + \frac{3}{2\mu} + \frac{3}{(\mu-1)} \right. \nonumber \\
&-& \left. \frac{1}{3(\mu-2)^2} - \frac{28}{9(2\mu-1)} - \frac{35}
{18(\mu-2)} \right]
\end{eqnarray}
which is the main result of this paper, and is an arbitrary dimension
expression for the $O(1/\Nf^2)$ part of the electron anomalous dimension in the
Landau gauge and as was noted earlier encodes information on the gauge
independent part of the $4$-dimensional renormalization group function,
$\gamma(g)$.

Aside from the internal consistency checks on the explicit derivation of
the values of the two loop integrals already discussed, we have checked that
the final result for $\eta_2$ agrees with the three loop $\overline{\mbox{MS}}$
anomalous dimension, $\gamma(g_c)$, expanded near four dimensions. It is worth
noting that this three loop result is in fact another very stringent check on
our result since we have only evaluated two $2$-loop graphs. Moreover, we can
now derive the coefficients of the higher order terms of $\gamma(g)$ at
$O(1/\Nf^2)$, which have not been given previously. For example, from (2.5),
with (2.3) and (7.4), we find
\begin{eqnarray}
\gamma(g) &=& - \, \frac{[4\Nf+3]}{16} g^2 + \frac{[40\Nf^2 + 54\Nf + 27]}{576}
g^3 \nonumber \\
&+& \left[ \frac{35\Nf^3}{1296} + \Nf^2 \left( \frac{1}{27} - \frac{\zeta(3)}
{4} \right) + c_1 \Nf + d_1 \right] g^4 + O(g^5)
\end{eqnarray}
where the unknown constants $c_1$ and $d_1$ can only be deduced from $\eta$ at
$O(1/\Nf^3)$ and $O(1/\Nf^4)$ respectively. The next to leading order
coefficients with respect to $\Nf$, at each subsequent perturbative order,
are deduced by first extracting all the $O(1/\Nf^2)$ corrections to $g_c$,
which are contained within the large $\Nf$ $\beta$-function of \cite{8}.
Comparing the numerical structure of (7.5) with the four loop $\beta$-function
of (2.2), we note that they are similar in that the transcendental number,
$\zeta(3)$, appears at fourth order but not at leading order in $\Nf$.

As (7.4) is valid in arbitrary dimensions, we can evaluate it in three
dimensions and find
\begin{equation}
\eta ~=~ - \, \frac{8}{3\pi^2\Nf} ~+~ \frac{16(32-3\pi^2)}{9\pi^4\Nf^2}
{}~+~ O\left( \frac{1}{\Nf^3} \right)
\end{equation}
where the $O(1/\Nf)$ term agrees with the calculation of \cite{16}. From (7.6),
we can obtain estimates for the anomalous dimension for various values of
$\Nf$. For instance, at $\Nf$ $=$ $3$, $\eta$ $=$ $-$ $0.09$ and at $\Nf$ $=$
$4$, $\eta$ $=$ $-$ $0.06$.

\sect{Discussion.}

We conclude with various observations. First, we have given an analytic
expression for the electron anomalous dimension at $O(1/\Nf^2)$ by solving the
appropriate Dyson equations at criticality within the large $\Nf$ expansion,
algebraically. Indeed, the information contained in the $d$-dimensional
expression, (7.4), relates to the perturbative coefficients of the gauge
independent part of the electron anomalous dimension. For completeness, we note
that this model is now solved at leading order in large $\Nf$, since the
critical exponents, or equivalently, the appropriate renormalization group
functions, relating to the $\beta$-function and the electron mass anomalous
dimension, $\gamma_m(g)$, have been given elsewhere, \cite{8}, and in the
notation of this paper, they are
\begin{eqnarray}
\gamma_m(g_c) &=& - \, \frac{2\eta_1}{(\mu-2)\Nf} \\
\lambda &=& (\mu-2) ~-~ \frac{(2\mu-3)(\mu-3)\eta_1}{\Nf}
\end{eqnarray}
where $2\lambda$ $=$ $-\, \beta^\prime(g_c)$. So, for example, in three
dimensions there is no $O(1/\Nf)$ correction to the $\beta$-function which is
gauge independent and this is consistent with the unltraviolet
superrenormalizability of that theory. Indeed the four dimensional perturbative
information, which is encoded within (7.4), (8.1) and (8.2), will provide
useful checks for future explicit perturbative calculations. Second, with the
techniques given in this paper, it ought now to be possible to go beyond the
$O(1/\Nf)$ expressions in (8.1) and (8.2), by extending our critical Dyson
equation approach. For instance, the $O(1/N^2)$ corrections to the
$\beta$-function of the bosonic and supersymmetric $O(N)$ $\sigma$ models are
known, \cite{2,9}. Indeed, the method of \cite{1,2} is such that one has always
to first compute the exponent $\eta$, which we have done here, before
attempting to determine $\lambda$ at the same order in large $N$ since, for
example, one needs to know $z_2$ which is determined from (3.9) and (3.13) once
$\eta_2$ is available. Finally, we remark that the methods which have been
developed here for computing anomalous dimensions in an abelian gauge theory,
will be applicable in solving similar models, such as the bosonic $CP(N)$
$\sigma$ model, beyond the $O(1/N)$ exponents which are presently known,
\cite{23}.

\vspace{1cm}
\noindent
{\bf Acknowledgements.} The author thanks Dr D.J. Broadhurst for providing a
copy of reference \cite{C} and for discussions on the gauge independence of
$\gamma(g)$ beyond one loop in QED, as well as Dr O.V. Tarasov for an indirect
communication giving the three loop structure of $\gamma(g)$. Drs D.R.T. Jones
and H. Osborn are also thanked for brief conversations.
\newpage
\appendix
\sect{Conventions.}
In this appendix, we briefly list our conventions and several results
involving $\gamma$-matrices which we required. Although we work in
arbitrary dimensions we take the trace convention to be $\mbox{tr}1$
$=$ $4$, and to manipulate $\gamma$-matrices we made use only of
\begin{equation}
\{ \gamma^\mu , \gamma^\nu \} ~=~ 2 \eta^{\mu \nu}
\end{equation}
working in Euclidean space throughout. From (A.1) we derive,
\begin{eqnarray}
\gamma^\sigma \gamma^\mu \gamma_\sigma &=& - \, 2(\mu-1) \gamma^\mu \\
\gamma^\sigma \gamma^\mu \gamma^\nu \gamma_\sigma &=& 4 \eta^{\mu \nu}
{}~+~ 2(\mu-2) \gamma^\mu \gamma^\nu \\
\gamma^\rho \gamma^\mu \gamma^\nu \gamma^\sigma \gamma_\rho &=&
4\eta^{\mu \sigma} \gamma^\nu - \, 2 \gamma^\mu \gamma^\sigma \gamma^\nu
\nonumber \\
&-& 2 \gamma^\nu \gamma^\mu \gamma^\sigma - \, 2(\mu-1) \gamma^\mu
\gamma^\nu \gamma^\sigma
\end{eqnarray}
Although we were able to manipulate the two loop integrals to reduce the
number of $\gamma$-matrices involved, we needed
\begin{equation}
\mbox{tr} (\gamma^\mu \gamma^\nu \gamma^\sigma \gamma^\rho) ~=~
4[ \eta^{\mu \nu} \eta^{\sigma \rho} \, - \, \eta^{\mu \sigma}
\eta^{\nu \rho} \, + \, \eta^{\mu \rho} \eta^{\nu \sigma}]
\end{equation}
to complete the calculations of $\Pi_{\mu \nu}$ and $\Sigma$.
\sect{Summary of recursion relations.}
To compute various basic bosonic two loop building block integrals making
up $\Sigma$ and $\Pi_{\mu \nu}$, we required various recursion relations,
which we list in this appendix. Whilst this list is perhaps not complete,
given the large number of tranformations one can make on the basic two
loop integral of fig. 10, they were all that we required for our purposes.
To compactify our expressions a little, we define a new five argument
quantity similar to
$\la$$\alpha_1,$$\alpha_2,$$\alpha_3,$$\alpha_4,$$\alpha_5$$\ra$, but
involving square brackets. In the following it will denote the two loop graph
obtained from the general graph of fig. 10, but with the respective exponents
adjusted by the arguments of the expression involving square brackets. For
instance, $[0,0,0,0,1]$ means
$\la$$\alpha_1,$$\alpha_2,$$\alpha_3,$$\alpha_4,$$\alpha_5$$+$$1$$\ra$ etc.
Also, when a square bracket appears with a subscript $\pm$ it will correspond
to multiplying the overall expression by a factor $(x^2)^{\mp 1}$, so that each
term has the correct dimensions. We also use the notation of \cite{2} and
define
\begin{eqnarray}
\alpha_1 + \alpha_2 + \alpha_5 ~=~ s_1 &,&
\alpha_3 + \alpha_4 + \alpha_5 ~=~ s_2 \nonumber \\
\alpha_1 + \alpha_4 + \alpha_5 ~=~ t_1 &,&
\alpha_2 + \alpha_3 + \alpha_5 ~=~ t_2 \nonumber \\
\alpha_1 + \alpha_2 + \alpha_3 + \alpha_4 + \alpha_5 ~=~ d &&
\end{eqnarray}
{}From the rule of fig. 12, we have derived the following recursion relations,
which we believe have not been given before, and emphasise that the relations
are completely general and applicable to any graph of the form of fig. 10,
and not solely for the specific problem dealt with in this paper. So, we have,
\begin{eqnarray}
\la \alpha_1, \alpha_2, \alpha_3, \alpha_4, \alpha_5 \ra
&=& \frac{(3\mu-d)(d-2\mu-1)}{(\alpha_2-1)(\alpha_3-1)} [0,-1,-1,0,1]_+
\nonumber \\
&&+ ~ \frac{(\alpha_2+\alpha_3-\mu-1)}{(\alpha_2-1)} [0,-1,0,0,1]
\nonumber \\
&&+ ~ \frac{(\alpha_2+\alpha_3 -\mu -1)}{(\alpha_3-1)} [0,0,-1,0,1] \\
&=& \frac{\alpha_1(\mu-\alpha_1-1)}{(\alpha_2-1)(\alpha_5-1)}
[1,-1,1,0,-1] \nonumber \\
&&+ ~ \frac{(\alpha_2+\alpha_5-\mu-1)}{(\alpha_2-1)}
[0,-1,1,0,0] \nonumber \\
&&+ ~ \frac{(\alpha_2+\alpha_5 -\mu-1)}{(\alpha_5-1)} [0,0,1,0,-1] \\
&=& \frac{\alpha_5(\mu-\alpha_5-1)}{(\alpha_1-1)(\alpha_2-1)} [-1,-1,0,0,1]_+
\nonumber \\
&&+ ~ \frac{(\alpha_1+\alpha_2-\mu-1)}{(\alpha_2-1)} [0,-1,0,0,0]_+ \nonumber
\\
&&+ ~ \frac{(\alpha_1+\alpha_2-\mu-1)}{(\alpha_1-1)} [-1,0,0,0,0]_+ \\
&=& \frac{\alpha_5(\mu-\alpha_2-\alpha_5-1)}{(2\mu-s_1-1)(s_1-\mu)}
[0,0,0,-1,1] \nonumber \\
&&+ ~ \frac{\alpha_2(\mu-\alpha_2-\alpha_5-1)}{(2\mu-s_1-1)(s_1-\mu)}
[0,1,0,0,0]_- \nonumber \\
&&+ ~ \frac{\alpha_2 \alpha_5}{(2\mu-s_1-1)(s_1-\mu)} [-1,1,-1,0,1] \\
&=& \frac{\alpha_3(2\mu-s_1)(s_1-\mu-1)}{(2\mu-s_2-1)(s_2-\mu)
(\alpha_1-1)} [-1,0,1,0,0] \nonumber \\
&&+ ~ \frac{\alpha_5(\alpha_1-\alpha_3-1)(\mu-\alpha_5-1)}
{(s_2-\mu)(2\mu-1-s_2)(\alpha_1-1)} [-1,0,0,0,1] \nonumber \\
&&+ ~ \frac{\alpha_3(\alpha_1-\alpha_3-1)}{(s_2-\mu)(2\mu-s_2-1)}
[0,0,1,0,0]_- \\
&=& \frac{\alpha_3(t_2-\mu)(\mu-\alpha_3-1)}{(d-2\mu)(3\mu-d-1)
(\alpha_5-1)} [0,1,1,0,-1]_- \nonumber \\
&&+ ~ \frac{\alpha_4(\mu-\alpha_2-\alpha_3-1)(\mu-\alpha_4-1)}{(d-2\mu)
(3\mu-d-1)(\alpha_5-1)} [0,1,0,1,-1]_- \nonumber \\
&&+ ~ \frac{(t_2-\mu)(\mu-\alpha_2-\alpha_3-1)}{(d-2\mu)(3\mu-d-1)}
[0,1,0,0,0]_- \\
&=& \frac{\alpha_2\alpha_3}{(2\mu-t_1)(t_1-\mu-1)} [0,1,1,0,-1]_-
\nonumber \\
&&+ ~ \frac{\alpha_3(\mu-\alpha_2-\alpha_3-1)}{(2\mu-t_1)
(t_1-\mu-1)} [0,0,1,-1,0] \nonumber \\
&&+ ~ \frac{\alpha_2(\mu-\alpha_2-\alpha_3-1)}{(2\mu-t_1)(t_1-\mu-1)}
[-1,1,0,0,0] \\
&=& \frac{(2\mu-s_1)(2\mu-s_2)}{(2\mu-t_2)(t_2-\mu-1)} [0,0,0,0,-1]_+
\nonumber \\
&&+ ~ \frac{(2\mu-s_2)(d+\alpha_5-3\mu-1)}{(2\mu-t_2)(t_2-\mu-1)}
[0,0,-1,0,0]_+ \nonumber \\
&&+ ~ \frac{(2\mu-s_1)(d+\alpha_5-3\mu-1)}{(2\mu-t_2)(t_2-\mu-1)}
[0,-1,0,0,0]_+ \\
&=& \frac{(2\mu-t_1-1)(t_1-\mu)}{(\alpha_2-1)(\alpha_3-1)} [0,-1,-1,0,1]_+
\nonumber \\
&&+ ~ \frac{(\alpha_2+\alpha_3-\mu-1)}{(\alpha_2-1)} [0,-1,0,-1,1]_+
\nonumber \\
&&+ ~ \frac{(\alpha_2+\alpha_3-\mu-1)}{(\alpha_3-1)} [-1,0,-1,0,1]_+ \\
&=& \frac{\alpha_3(t_2-\mu)(2\mu-1-t_2)}{(\alpha_1-1)(t_1-\mu-1)
(2\mu-t_1)} [-1,0,1,0,0] \nonumber \\
&&+ \, \frac{(\alpha_1-\alpha_3-1)(3\mu-d)(d-2\mu-1)}{(\alpha_1-1)
(t_1-\mu-1)(2\mu-t_1)} [-1,0,0,0,0]_+ \nonumber \\
&&+ ~ \frac{\alpha_3(\alpha_1-\alpha_3-1)}{(2\mu-t_1)(t_1-\mu-1)}
[0,0,1,0,-1]
\end{eqnarray}
Further, we record the recursion relation derived using integration
by parts which we also required in evaluating $\Pi_{\mu\nu}$, which is,
\cite{2,18},
\begin{eqnarray}
\frac{\la \alpha_1, \alpha_2, \alpha_3, \alpha_4, \alpha_5 \ra}
{(d+t_1-4\mu)^{-1}}
&=& \alpha_2 ( [0,1,0,0,0]_- - [-1,1,0,0,0])
\nonumber \\
&+& \alpha_3 ( [0,0,1,0,0]_- - [0,0,1,-1,0] )
\end{eqnarray}
\sect{Basic two loop integrals.}
In this appendix, we give a list of the basic bosonic two loop integrals
required to compute $\Sigma$, $\Pi$ and $\Xi$, expanded to the finite part
with respect to $\Delta$. As discussed in sect. 5 these fall into classes ie
those which are computed directly using the method of subtractions and
those which are not but which are determined by using recursion relations.
As the former set are easy to establish, we list only those of the second
class which we require. First, we consider the basic graphs for $\Pi$ and
$\Xi$, using the notation of fig. 10. We note that the integral
$\la \alpha, \alpha, \alpha, \alpha, \beta-2 -\Delta \ra$ has already been
discussed earlier.
\begin{equation}
\la \alpha, \alpha-1, \alpha-1, \alpha, \beta-\Delta \ra ~=~
- \, \frac{2\pi^{2\mu}(\mu-1)a(\alpha)a(\alpha-1)a(\beta)}{\Gamma(\mu+1)}
\end{equation}
\begin{eqnarray}
\la \alpha-1, \alpha, \alpha, \alpha, \beta-\Delta \ra
&=& \frac{2 \pi^{2\mu} a(\alpha) a(\alpha-1) a(\beta)}{\Gamma(\mu)}
\left[ \frac{1}{\Delta} + B(\beta) \right.
\nonumber \\
&-& \left. B(\alpha-1) - \frac{(\beta-2)}{(\beta-1)}
- \frac{(\mu + \beta - 1)}{2\mu(\alpha-1)} \right]
\end{eqnarray}
\begin{eqnarray}
\la \alpha-1, \alpha-1, \alpha, \alpha, \beta-\Delta \ra
&=& \frac{2 \pi^{2\mu} a(\alpha-1) a(\alpha) a(\beta)}{\Gamma(\mu)}
\left[ \frac{1}{\Delta} + B(\beta) \right.
\nonumber \\
&-& \left. B(\alpha-1) - \frac{1}{\alpha-1} + \frac{2}{\mu-1} \right]
\end{eqnarray}
\begin{eqnarray}
\la \alpha, \alpha-1, \alpha-1, \alpha, \beta-1-\Delta \ra &=&
\!\!\! \frac{\pi^{2\mu} a(\beta-1)}{a^2(\mu-\alpha) \Gamma(\mu)} \!
\left[\frac{1}{\Delta} + B(\beta) - B(\alpha-1) \right. \nonumber \\
&+& \left. \frac{1}{\mu-1} \right]
+ \frac{(\beta+1-\mu)(2\alpha-3)}{(\alpha-1)^2} \nonumber \\
&\times& \!\!\!\!\! \la \alpha-1, \alpha-1, \alpha-1, \alpha-1, \beta \ra
\end{eqnarray}
\begin{eqnarray}
\la \alpha-1, \alpha, \alpha-1, \alpha, \beta-1-\Delta \ra &=&
\frac{\pi^{2\mu}(\mu-1) a^2(\alpha)a(\beta)}{\Gamma(\mu)} \nonumber \\
&+& \frac{(\mu-1-\beta)(2\alpha-3)}{(\alpha-1)^2} \nonumber \\
&\times& \!\!\!\!\! \la \alpha-1, \alpha-1, \alpha-1, \alpha-1, \beta \ra
\end{eqnarray}
\begin{eqnarray}
\la \alpha, \alpha, \alpha-2, \alpha, \beta-\Delta \ra &=&
\frac{\pi^{2\mu} a^2(\alpha) a(\beta-1)}{\Gamma(\mu+1)}
\left[ \frac{1}{\Delta} + B(\beta) - B(\alpha-1) \right. \nonumber \\
&+& \left. \frac{\mu^2 + \mu -1}{\mu(\mu-1)}
- \frac{3}{2} - \frac{(\alpha-1)^2}{(\mu-\beta)} \right. \nonumber \\
&+& \left. \frac{\alpha}{\beta-1}
- \frac{\alpha(\alpha-2)(\beta-1)}{2(\alpha-1)(\mu-\beta)} \right]
\end{eqnarray}
\begin{eqnarray}
\la \alpha, \alpha-1, \alpha-2, \alpha, \beta-\Delta \ra &=&
\frac{\pi^{2\mu}(2-\mu) a(\alpha) a(\alpha-1) a(\beta)}{\Gamma(\mu+1)}
\left[ \frac{1}{\Delta} + B(\beta) \right. \nonumber \\
&-& \left. B(\alpha-1) + \frac{\mu^2 + \mu -1}{\mu(\mu-1)}
+ \frac{\mu-1}{\mu-2} \right. \nonumber \\
&-& \left. \frac{(\beta-1)}{(\mu-2)(\alpha-1)} \right]
\end{eqnarray}
Next, the basic integrals for $\Sigma$ are,
\begin{equation}
\la \alpha, \beta-1-\Delta , \alpha-1, \beta-\Delta, \alpha \ra ~=~
\frac{\pi^{2\mu} a(\alpha-1) a(\alpha) a(\beta) (\mu-1)}{\Gamma(\mu+1)}
\left[ \frac{\beta-1}{\alpha-1} - 3 \right]
\end{equation}
\begin{eqnarray}
\la \alpha-1, \beta-\Delta, \alpha, \beta-\Delta, \alpha \ra &=&
\frac{\pi^{2\mu} a(\alpha-1) a(\alpha) a(\beta)}{\Gamma(\mu)}
\left[ \frac{2}{\Delta} + B(\beta) - 3 \right.
\nonumber \\
&-& \left. B(\alpha-1) + \frac{\beta-1}{\alpha-1} - \frac{1}{\mu} \right]
\end{eqnarray}
\begin{eqnarray}
\la \alpha, \beta-\Delta, \alpha, \beta-1-\Delta, \alpha \ra &=&
\frac{\pi^{2\mu} a^2(\alpha) a(\beta-1)}{\Gamma(\mu)}
\left[ \frac{2}{\Delta} + B(\beta) - B(\alpha-1) \right. \nonumber \\
&+& \left. \frac{1}{\beta-1} + \frac{\alpha(\beta-1)}{2(\alpha-1)(\mu-\beta)}
\right. \nonumber \\
&-& \left. \frac{(3\mu-1)(\alpha-1)}{2\mu(\mu-\beta)} - \frac{2}{\mu} \right]
\end{eqnarray}
\begin{eqnarray}
\la \alpha-1, \beta+1-\Delta, \alpha, \beta-\Delta, \alpha-1 \ra &=&
\frac{\pi^{2\mu} a^2(\alpha-1) a(\beta+1)}{(\mu-\alpha)
\mu\Gamma(\mu-1)} \left[ \mu-\alpha \right. \nonumber \\
&+& 1 \left. - \, \frac{\beta(\mu-\beta+\alpha-2)} {(\alpha-1)} \right]
\end{eqnarray}
\begin{eqnarray}
\la \alpha-1, \beta-1-\Delta, \alpha, \beta-\Delta, \alpha \ra &=&
\frac{\pi^{2\mu} a(\alpha-1) a(\alpha) a(\beta)}{\Delta \Gamma(\mu)}
\left[ \left( 3 - \frac{(\beta-1)}{(\alpha-1)} \right) \right. \nonumber \\
&+& \left. \Delta \left( B(\beta) - B(\alpha-1) + \frac{1}{\mu-1}
\right) \right]
\end{eqnarray}
\begin{eqnarray}
\la \alpha, \beta-2-\Delta, \alpha-1, \beta-\Delta, \alpha \ra &=&
\frac{\pi^{2\mu} (2-\mu) a(\alpha-1) a(\alpha) a(\beta)}{\Gamma(\mu+1)}
\nonumber \\
&\times& \!\! \left[ \frac{1}{\Delta} + B(\beta) - B(\alpha-1) \right.
\nonumber \\
&+& \left. \frac{\mu^2 + \mu -1}{\mu(\mu-1)} + \frac{2(\mu-\beta)}{(\alpha-1)}
\right]
\end{eqnarray}
\begin{eqnarray}
\la \alpha, \beta-2-\Delta, \alpha, \beta-\Delta, \alpha \ra &=&
\frac{2 \pi^{2\mu} a(\alpha-1) a(\alpha) a(\beta)}{(\beta-1) \Gamma(\mu)}
\nonumber \\
&+& \frac{\pi^{2\mu} a(\alpha) a(\alpha-1) a(\beta)}{\Gamma(\mu)}
\left[ \frac{(\beta-1)(\mu-1)}{\mu(\alpha-1)^2} \right. \nonumber \\
&&- \left. \frac{1}{(\alpha-1)} - \frac{4(\mu-\beta)}
{(\alpha-1)}\right. \nonumber \\
&+& \left. \left( 1 - \frac{(\beta-1)(\mu-1)}{\mu(\alpha-1)} \right)
\left( \frac{1}{\Delta} +  B(\beta) \right. \right. \nonumber \\
&-&  \left. \left. B(\alpha-1) + \frac{\mu^2 + \mu -1} {\mu(\mu-1)}
\right) \right]
\end{eqnarray}
\begin{eqnarray}
\la \alpha+1, \beta-2-\Delta, \alpha, \beta-1-\Delta, \alpha \ra &=&
\frac{\pi^{2\mu} a^2(\alpha) a(\beta-1) (\beta-1)}{2(\mu-\alpha
-1) \Gamma(\mu+1)} \nonumber \\
&\times& \left[ \frac{(\beta-2)(\beta-3)}{2\alpha} + \mu-\beta+1
\right. \nonumber \\
&&- \left. \frac{4(\alpha-1)(\beta-2)}{(\beta-1)} \right]
\end{eqnarray}
\begin{eqnarray}
\la \alpha+1, \beta-1-\Delta, \alpha, \beta-\Delta, \alpha-1 \ra &=&
\frac{\pi^{2\mu} a(\alpha) a(\alpha-1) a(\beta-1)}
{(\mu-\alpha-1) \Gamma(\mu+1)} \nonumber \\
&\times& \!\!\!\left[ \frac{\mu-\alpha+1}{\beta - 1} - \frac{(\mu-1)}{\alpha}
\right]
\end{eqnarray}
\begin{eqnarray}
&&\la \alpha+1, \beta-1-\Delta, \alpha, \beta-1-\Delta, \alpha \ra \nonumber \\
&&= \frac{\pi^{2\mu} a(\alpha) a(\alpha-1) a(\beta-1)}{\Gamma(\mu)
(\alpha-1)(\beta-1)} \left[ \frac{4}{\Delta} + \frac{2}{\beta-1}
- \frac{2}{\alpha} + \frac{2}{\mu-\alpha-1} \right. \nonumber \\
&&+ \left. 2(B(\beta) - B(\alpha-1)) + \frac{(\beta-1)}
{(\mu-\beta)(\alpha-1)} \right. \nonumber \\
&&+ \left. \frac{4(2\alpha+2\beta-3)(\alpha-\mu-1)(\alpha-1)}
{\mu(\mu-1)(\mu-\alpha-1)} + \frac{4(\mu+\alpha-3)}{(\mu-1)(\mu-\alpha-1)}
\right. \nonumber \\
&&- \left. \frac{2(\alpha+\beta-1)(\alpha-\mu-1)(\beta-2)}
{\mu(\mu-1)(\mu-\alpha-1)} \left( \frac{\beta-3}{2\alpha}
+ \frac{\mu-\beta+1}{\beta-2} \right) \right. \nonumber \\
&&- \left. \frac{(\beta-1)^2(\mu-\beta)(\mu-\alpha+1)}
{\alpha\mu(\mu-1)(\mu-\alpha-1)} \left( \frac{\alpha}{\beta-1}
- \frac{(\mu-1)}{(\mu-\alpha+1)} \right) \right]
\end{eqnarray}

\newpage
\noindent
{\Large {\bf Figure Captions.}}
\begin{description}
\item[Fig. 1.] Skeleton Dyson equation for the electron.
\item[Fig. 2.] Skeleton Dyson equation for the photon.
\item[Fig. 3.] Uniqueness rule for a bosonic vertex.
\item[Fig. 4.] Electron photon vertex.
\item[Fig. 5.] Integration by parts rule for gauge vertex.
\item[Fig. 6.] Additional integration by parts rule.
\item[Fig. 7.] Further integration by parts rule for gauge vertex.
\item[Fig. 8.] Graphical representation of $\Pi_{\mu \nu}(x)$.
\item[Fig. 9.] Photon self energy after integrating by parts.
\item[Fig. 10.] Basic two loop self energy graph.
\item[Fig. 11.] Various basic bosonic graphs contributing to $\Pi_{\mu \nu}$.
\item[Fig. 12.] Basic rule for recursion relations.
\end{description}

\newpage
\begin{thebibliography}{99}
\bibitem{1} A.N. Vasil'ev, Yu.M. Pis'mak \& J.R. Honkonen, Theor. Math.
Phys. {\bf 46} (1981), 157.
\bibitem{2} A.N. Vasil'ev, Yu.M. Pis'mak \& J.R. Honkonen, Theor. Math.
Phys. {\bf 47} (1981), 291.
\bibitem{3} J.A. Gracey, J. Phys. {\bf A24} (1991), L431.
\bibitem{4} J.A. Gracey, Int. J. Mod. Phys. {\bf A6} (1990), 395.
\bibitem{5} J.A. Gracey, Nucl. Phys. {\bf B348} (1991), 737.
\bibitem{6} D. Nash, Phys. Rev. Lett. {\bf 62} (1989), 3024.
\bibitem{7} D. Espriu, A. Palanques-Mestre, P. Pascual \& R. Tarrach, Z.
Phys, {\bf C13} (1982), 153.
\bibitem{8} A. Palanques-Mestre \& P. Pascual, Comm. Math. Phys. {\bf 95}
(1984), 277.
\bibitem{9} J.A. Gracey, Nucl. Phys. {\bf B352} (1991), 183.
\bibitem{10} S.G. Gorishny, A.L. Kataev \& S.A. Larin, Phys. Lett. {\bf
194B} (1987), 429; S.G. Gorishny, A.L. Kataev, S.A. Larin \& S.R.
Surguladze, Proc. Quark '90 Conference (Telavi, 1990) eds A.N. Tavkhelidze,
V.A. Matveev, V.A. Rubakov \& P.G. Tinyakov (World Scientific, Singapore,
1991); Phys. Lett. {\bf 256B} (1991), 81.
\bibitem{11} P. Pascual \& R. Tarrach, `QCD: renormalization for the
practitioner' Lecture Notes in Physics {\bf 194} (Berlin, Springer, 1984).
\bibitem{A} A. Palanques-Mestre \& P. Pascual, Commun. Math. Phys. {\bf 95}
(1984), 277; D. Espriu, A. Palanques-Mestre, P. Pascual \& R. Tarrach, Z. Phys.
{\bf C13} (1982), 153.
\bibitem{B} J.A. Gracey, Int. J. Mod. Phys. {\bf A8} (1993), 2465.
\bibitem{12} E.S. Egoryan \& O.V. Tarasov, Theor. Math. Phys. {\bf 41}
(1979), 26.
\bibitem{C} O.V. Tarasov, `Anomalous dimensions of quark masses in three loop
approximation', JINR preprint P2-82-900, (in Russian).
\bibitem{D} B. Zumino, J. Math. Phys. {\bf 1} (1960), 1; N.N. Boguliobov \&
D.V. Shirkov, `Quantum fields', (Benjamin Cummings, Reading, Mass, 1983).
\bibitem{13} D.J. Amit, `Field theory, the renormalization group and
critical phenomena' (McGraw-Hill, New York, 1978).
\bibitem{14} A.N. Vasil'ev, Yu.M. Pis'mak \& J.R. Honkonen, Theor. Math.
Phys. {\bf 48} (1981), 750.
\bibitem{15} A.A. Slavnov \& L.D. Faddeev, `Introduction to the quantum
theory of gauge fields' (Nauka, Moscow, 1978).
\bibitem{16} G.W. Semenoff \& L.C.R. Wijewardhana, Phys. Rev. {\bf D45}
(1992), 1342.
\bibitem{17} M. d'Eramo, L. Peliti \& G. Parisi, Lett. Nuovo Cim. {\bf 2}
(1971), 878.
\bibitem{18} D.I. Kazakov, Phys. Lett. {\bf 133B} (1983), 606; Theor.
Math. Phys. {\bf 58} (1985), 343; Theor. Math. Phys. {\bf 62} (1985), 127.
\bibitem{19} N.I. Ussyukina, Theor. Math. Phys. {\bf 54} (1983), 78; Phys.
Lett. {\bf 267B} (1991), 382.
\bibitem{20} D.I. Kazakov \& A.V. Kotikov, Theor. Math. Phys. {\bf 73} (1987),
1264.
\bibitem{21} J.A. Gracey, Phys. Lett. {\bf 277B} (1992), 469.
\bibitem{22} J.A. Gracey, J. Phys. {\bf A25} (1992), L109.
\bibitem{23} A.N. Vasil'ev \& M.Yu. Nalimov, Theor. Math. Phys. {\bf 56}
(1983), 643.
\bibitem{28} K.G. Chetyrkin, A.L. Kataev \& F.V. Tkachov, Nucl. Phys. {\bf
B174} (1980), 345.
\end{thebibliography}
\end{document}